\documentstyle[prl,aps,multicol,epsfig,tabularx]{revtex} 
\def\xv{{\bf x}}

\begin{document} 

\draft 
 
\title{Extinction transition in bacterial colonies under forced 
convection} 
\author{T. Neicu,$^{1}$ A. Pradhan,$^{1}$ D. A.
Larochelle,$^{2}$  and A. Kudrolli$^{1,\dag}$} 
\address{$^{1}$Department of Physics, Clark University, Worcester, 
Massachusetts 
01610 \\
$^{2}$Department of Biology, Clark University, Worcester,
Massachusetts 01610} 

\date{October 15, 1999} \maketitle 
 
\begin{abstract} 
 
We report the spatio-temporal response of {\it Bacillus subtilis} growing 
on a nutrient-rich layer of agar to ultra-violet (UV) radiation. Below a 
crossover temperature, the bacteria are confined to regions that are 
shielded from UV radiation. A forced convection of the
population is effected by rotating a UV radiation shield relative to 
the petri dish. The extinction speed at which the bacterial 
colony lags behind the
shield is found to be qualitatively similar to the front velocity
of the colony growing in the absence of the hostile environment as 
predicted by the model of Dahmen, Nelson and Shnerb. A quantitative
comparison is not possible without considering the slow dynamics
and the time-dependent interaction of the population with the
hostile environment.

\end{abstract} 
 
 


Bacterial colonies growing on a nutrient rich substrate have served
as  model systems for studying pattern formation and population
dynamics in biological systems. Studies with strains of {\it
Bacillus subtilis} and {\it Escherichia coli} have reported a wide
variety of complex patterns depending on nutrient conditions 
\cite{murray91,matsushita90,benjacobs94,budrene95}. The
patterns have been modeled using reaction-diffusion 
equations~\cite{brenner98,kozlovsky99,lacasta99}. These
experimental and theoretical studies have considered an
essentially uniform environment where the  changes are due only to
the depletion of nutrients with time.  However, living organisms
often are forced to migrate due to changes in the environment.

The
modeling of population dynamics of bacterial colonies due to
changes in the environment has been studied recently by Shnerb,
Nelson, and Dahmen~\cite{shnerb98,dahmen98}.
Their theoretical model incorporates
the effect of a forced convection on the growth of a bacterial
colony by considering the convective-diffusion equation given by:
\begin{equation}
\frac{\partial c(\textbf{x},t)}{\partial t} = D \nabla^2 
c(\textbf{x},t) - 
\textbf{v} \cdot \nabla c(\textbf{x},t) + 
U(\textbf{x})c(\textbf{x},t)
- bc^2(\textbf{x},t)\,,
\label{genfisher}
\end{equation}
where $c(\textbf{x},t)$ is the bacteria number density, $D$ is the 
diffusion constant of the bacteria, $U(\textbf{x})$ is the
spatially varying growth  potential, $\textbf{v}$ is an externally
imposed convection velocity, and
$b$ is a parameter that limits the population number density to a
maximum saturation value. If $\textbf{v} = 0$ and 
$U(\textbf{x})$ is constant, Eq.~(\ref{genfisher}) corresponds to
the Fisher wave equation~\cite{fisher37} which has a
solution with a limiting constant value of the front speed
$v_{F}$. Wakita {\em et al.}~\cite{wakita94} have studied 
a colony of {\it Bacillus subtilis} in 
a high nutrient and low agar medium growing in such a Fisher mode.

The two new features of the forced convection model given by
Eq.~(\ref{genfisher}) are the introduction of a growth potential
$U(\textbf{x})$, corresponding to exposing  photosensitive bacteria
to a light source for example, and the convection of the
bacteria due to the motion $\textbf{v}$ of the light source. By
considering a colony confined to a rectangular region, the 
resulting steady-state number density of the bacteria (the time
independent  solution of Eq.~(\ref{genfisher})) was obtained
in Ref.~\cite{dahmen98}  as a function of
$\textbf{v}$. They concluded that that the total number of bacteria in 
the rectangular region decreases linearly to zero as $v$ approches 
$v_F$ from below. The steady-state spatial density distribution was obtained
by solving for the time-independent solutions of
Eq.~(\ref{genfisher}) numerically. Because the linearized version of 
Eq.~(\ref{genfisher})
allows a mapping to non-Hermitian quantum mechanics, additional
predictions of the properties of bacterial colonies in terms of 
localization-delocalization transitions in quantum systems can be
made~\cite{shnerb98}.

We report the first experimental study of a {\it Bacillus subtilis} 
colony forced to migrate by environmental changes due to
a moving ultra-violet (UV) source. UV radiation is shined
on a petri dish containing nutrient rich agar except in a
rectangular region which is shielded. Although UV radiation is
supposed to kill these bacteria~\cite{munakata81}, we find more
subtle behaviors. For example, the colony is confined to the shielded
region only when the temperature is below a ``crossover'' value of
approximately $22^\circ$C. When the UV radiation is turned off, the
front of the colony which was near the boundary between the hostile
and favorable regions initially grows slowly, but recovers to the
Fisher front speed $v_F$ in about 25 hours (h). This slow
recovery near the boundary suggests the presence of
signalling between the bacteria, a feature which is absent in
Eq.~(\ref{genfisher}).

To study the effect of a changing environment, we rotate the 
rectangular shield with a constant angular velocity relative to the
petri dish. The bacteria are inoculated along a line inside the
rectangular shield region. The rotation results in the colony being
forced to convect with velocities that increase linearly from zero
at the axis of rotation to a maximum value at the edge of the plate.
We find that the bacteria colony cannot keep up with the shielded
region if the shield moves with velocities much greater than
$v_{F}$, thus showing an extinction transition in qualitative
agreement with the theoretical model~\cite{dahmen98}. The spatial
number density
$n(\xv, v)$ of the bacteria as a function of the speed $v$ was
measured and found to be time-dependent, even after three days of
forced convection. These experimental results illustrate the
relevance of Eq.~(\ref{genfisher}) and also indicate that the
time-dependent response is of experimental relevance because of the
long time scales of biological systems.

We now describe our experimental procedure and observations in more
detail. The wild-type strain of {\it Bacillus subtilis} was obtained
from Presque  Isle Cultures and freeze dried at $-70^\circ$C. All
experiments  were performed from this initial sample by incubating
a portion  of the sample for 8 h at $30^\circ$C in nutrient rich
broth. A drop of this  broth representing a total of at least
$10^{7}$ bacteria is used to inoculate the nutrient rich agar. The
experiments were performed  in 15\,cm diameter plexiglass petri
dishes containing a thin layer of nutrient agar (7\,grams/liter of
bacto-peptone and 3\,grams/liter agar.) These conditions are
similar to that used in previous observations of the Fisher wave
mode~\cite{wakita94}. When inoculated as a single  point source
(diameter $\sim$ 3 mm), the growth of the colony was observed to
have a uniform disk shape with a front velocity that increases
slowly for the first 8 h and eventually reaches a constant
front speed $v_{F}$ consistent with previous work~\cite{wakita94}.
Experiments were performed over a range of temperature ($21^\circ -
40^\circ$C) and it was found that $v_F$ is an increasing function
of temperature within this range with
$v_F=1.7\,\mu {\rm m/s}$ at $40^\circ$C and
$v_F=0.19\,\mu {\rm m/s}$ at $21^\circ$C.

Next we describe our experiments in which we shine UV radiation on 
the petri dish using two 8\,W long wavelength UV-lamps placed 5\,cm
above the dish. An Aluminum sheet is used to shield a rectangular
region of the petri dish from the radiation (see Fig.~\ref{fisher}). The
density of the bacteria is obtained by imaging the light scattered by 
the bacteria with a
CCD camera. Calibration experiments show that the light intensity
is  proportional to the bacteria density. The colony at time
$t = 23.15$\,h after a point inoculation at the center of the
petri dish is shown in Fig.~\ref{fisher}a for $21^\circ$C. The 
shielded region is within the dashed lines and has
a width $w= 5$\,cm. We observe that the front of the colony is
circular and its diameter is smaller than $w$. As the colony grows
further outward, the edge of the shielded region  is reached, and
the shape of the colony is no longer circular as shown in
Fig.~\ref{fisher}b. The width of the colony along the axes parallel
($x$) and perpendicular ($y$) to the shield is plotted in
Fig.~\ref{fisher}c.  The error bars correspond to
the range of fluctuations due to slightly different initial
conditions in different  runs. The diameter $d$ of the bacterial
colony growing at the same temperature in a petri dish which is
completely shielded from UV radiation is also shown. We  observe
that the colony under the shield grows  with a speed comparable to
$v_{F}$ at that temperature. As the colony approaches the edge of
the shield, the front speed slows down because the bacteria are
confined.

To further demonstrate that the confinement effect is due to the
presence of UV radiation, the UV radiation was turned off after 72
h. We observe no change in the velocity of the front along
the  $x$-direction as expected, because the bacteria are deep
inside the shield. However, we would expect to see a change in the
rate of growth along the $y$-direction because the radiation has
been removed. We observe that the front velocity recovers to
$v_{F}$, but only after 25 h. This behavior is not modeled by
Eq.~(\ref{genfisher}), but is important in our discussion of the
convection experiments as discussed below.

We performed experiments at higher temperatures and observed that
for temperatures greater than approximately $22^\circ$C, the
bacteria are able to grow into irradiated regions, but with a front
speed that decreases with time. (At
$26^\circ$C, the speed was reduced by 41\% after 12 h.) Hence,
in the presence of radiation we can vary the growth rate by changing
the temperature and obtain a transition from a localized
colony to one which is delocalized. A detailed study of this
phenomena would be an interesting avenue for further research. In
this paper we will consider a simple case in which the
bacteria are confined at a temperature of $22\pm 1^\circ$C to
investigate the extinction transition in the presence of convection.

The convection experiments were performed by inoculating the 
bacteria along a diameter of the petri dish. The petri dish is then
kept under a radiation shield of width $w=4.3\,{\rm cm}$  and
placed  on a rotating platform, similar to the experiments
described earlier. As the platform rotates, the region shielded
from the UV radiation advances at a speed which increases linearly
from the axis of rotation outward. The colony was initially allowed to grow
for 14 h before the platform was rotated. During this time the
bacteria covered the shielded region. The time $t=0$ corresponds to
the time at which the platform was rotated. The results of the
colony growth under these conditions are shown in
Fig.~\ref{rot-pic}. The position of the shielded  region and the
axis of rotation are indicated. The  bacterial population is
clearly seen to migrate and follow the shielded region  at low
velocities near the axis of rotation and lag behind at higher 
velocities. 
 
These observations are consistent with the theory of
Ref.~\cite{dahmen98} where a phase diagram for the growth and the
extinction of a colony as a function of the growth potential
$U$ and the convection speed $v$  was
obtained using Eq.~(\ref{genfisher}). In particular, it was
predicted that the bacteria will be localized to the favorable
region. Furthermore, the total bacterial population in
favorable regions decreases linearly to zero as a function of $v$
as $v$ approaches  $v_{F}$ from below. (In this theory the 
critical extinction speed
$v_c$ is the same as $v_F$.) To make quantitative comparisons, we
have extracted the positions of the fronts corresponding to the
three images shown in Fig.~\ref{rot-pic}. These positions are
plotted in  Fig.~\ref{front-pic}a; the origin corresponds to the
axis of  rotation and the initial line of inoculation is along the
horizontal axis. The dashed
arc in Fig.~\ref{front-pic}a corresponds to the distance
where the velocity of the shield is the same as
$v_{F}$. We observe that very far from the axis of rotation, the
front does not change during the time $t = 46.56$ h to $t =
73.73$ h, indicating that bacteria which could not cope with
the speed of the shield were left behind in the hostile irradiated
region and did not grow.

Dividing the displacement of the bacteria front by the time
difference between images, we extracted the approximate velocity of the
front as a function of the radial distance $r$. Such an analysis
ignores the diffusion of the bacteria along the radial  direction.
The data for the average velocity of the front, $v_b(r)$, is plotted
in  Fig.~\ref{front-pic}b. The velocity of the shield $v(r)$ also
is  plotted to provide a reference for the front velocities. The
bacteria are confined to the shielded region, and $v_{b}(r)$ is
observed to increase with $r$, but is always less than $v(r)$, the
corresponding speed of the shield.  The reason for this lag might
be due to the slow recovery of the bacteria after the UV irradiated
region moves ahead as discussed earlier in reference to
Fig.~\ref{fisher}c. We also observe that $v_b(r)$ increases
linearly up to a velocity of
$0.2\,\mu {\rm m/s}$ which corresponds to $r \sim 45$\,mm. For
greater $r$,
$v_{b}(r)$ decreases and the bacteria increasingly lag behind the
shield and stop growing for $r > 80$\,mm. The maximum value of
$v_{b}$ corresponds to the value of
$v_{F}$ of the bacteria colony at $22^\circ$C in the absence of
convection and UV radiation.

To explain the velocity data for $r > 50$\,mm, we note  the
following. In the interval of time corresponding to the images
shown in Figs.~\ref{rot-pic}b and \ref{rot-pic}c, the point where
the bacteria completely lag behind the shield decreases from $r =
75$\,mm to $r = 59$\,mm. During this time, the bacteria are exposed
to UV radiation for at least part of this time interval which increases for
larger $r$. Hence, because the bacteria grow for only a portion of
the total time, the mean front speed $v_b(r)$ decreases. The front
speed is zero when the bacteria are always in the UV radiation
corresponding to $r > 80$\,mm in Fig.~\ref{front-pic}b.

We also note that because of the slow rate of growth of the
colony, the relative slow speed of the shield $v$, and the finite
width of the shield $w$, a long transient time of the order of $w/(v
- v_{c})$ is required for the shield to leave the colony which is
growing with a speed $v$ less than the critical extinction speed
$v_{c}$. This transient time diverges as $v$ approaches $v_{c}$.
Hence, for an experiment which is conducted over a finite duration,
the value of $r$ where the bacteria completely  lag behind the
shield is larger than the value corresponding to the critical
extinction speed $v_{c}$. However, 
$v_{c}$  can be indirectly calculated from the above relation for
the transient time. We obtain the estimate $v_{c}
\sim 0.23\,\mu {\rm m/s}$ which is similar to the value of 
$v_{F} \sim 0.26$ at $22^\circ$C. This estimate was obtained from the image
in Fig.~\ref{front-pic}  using
$v = 0.4\,\mu{\rm m/s}$ at $r = 59$\,mm where the bacteria have
completely lagged behind the shield  at time
$t= 73.73\,{\rm h}$ of rotation. Therefore, we find a
critical extinction speed consistent with the Fisher wave velocity
as predicted in Ref.~\cite{dahmen98}.

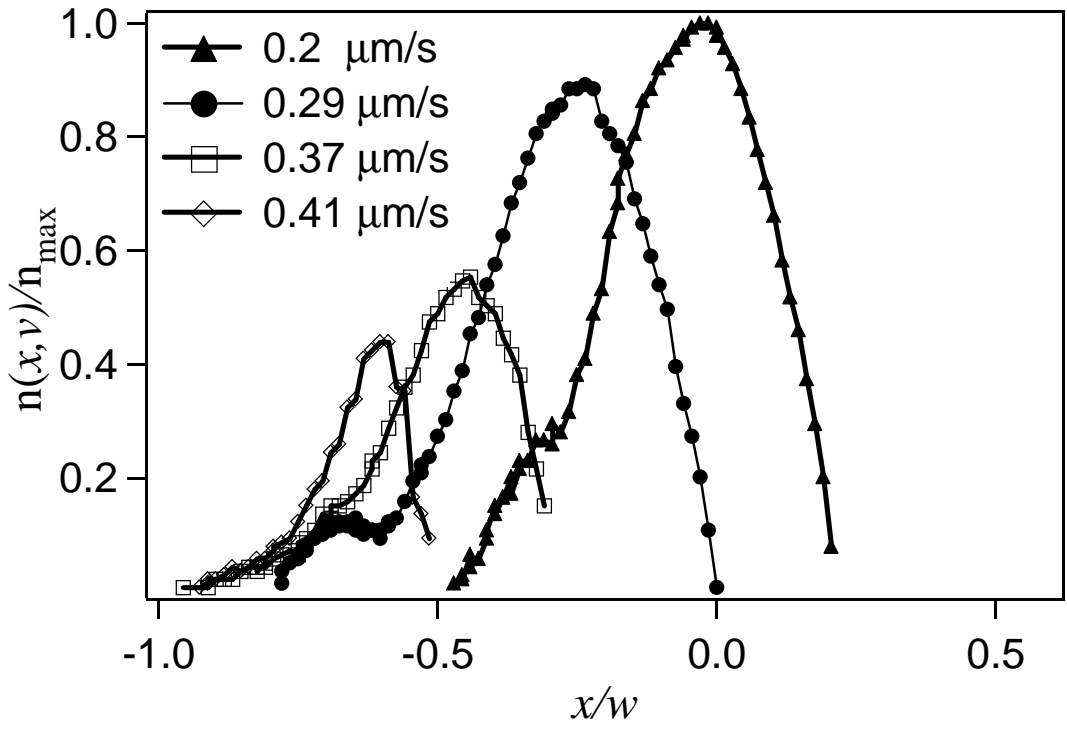
\begin{figure}
\begin{center}
\end{center} 
\caption{The number density $n(x,v)$ of bacteria at 
various speeds $v$ at $t = 73.73$ h normalized by the maximum number 
density $n_{\rm max}$. The horizontal axis is normalized with the
width 
$w$ of the shielded region and corresponds to 0.5 to $-0.5$. The
density  of the population is observed to decay to zero for higher
velocities but the  distribution is time-dependent.}
\label{transition}
\end{figure}

A more direct comparison of our experimental results to theory can
perhaps be made by considering the time dependent response of the
model considered in Eq.~(\ref{genfisher}). Additional considerations
such as the time-dependent response of the front speed of the bacteria
may have to be incorporated. To encourage future comparisons of
experimental data with time-dependent models, we plot the  number
density $n(x,v)$ of the bacteria colony at different distances 
from the shield in Fig.~\ref{transition}  corresponding to
different convection velocities $v$. The shielded  region
normalized by the width $w$ corresponds to $-0.5$ to 0.5. This data
corresponds to the image shown in Fig.~\ref{rot-pic}c. These
density distributions are still time 
dependent except at $v = 0.41\,\mu{\rm m/s}$, which corresponds to 
distances where the bacteria are immobile because they have been in
the UV irradiated regions for a long time. We observe  that the
front of the colony in the direction of the convection velocity
always lags behind the edge of the strip. This characteristic  of
the bacteria distribution is similar to that predicted in 
Ref.~\cite{dahmen98},
but a direct comparison is not possible because the distribution
is still time-dependent after $t=73$ h of rotation.  From
Fig.~\ref{transition} we further observe that the  total bacterial
population given by the area under the curve  decreases for increasing 
velocity. We have
found it impractical to conduct the experiments for a longer  time,
which is a significant limitation in making  a more direct
comparison with time-independent predictions.

The fact that the extinction transition occurs near $v_{F}$ is an 
interesting result for real biological systems
because of the  relatively simple model considered in
Refs.~\cite{shnerb98,dahmen98}. Our experiments are an important
first step in investigating the usefulness of convection-diffusion
models in studying convection in biological systems. The question
remains if the observed evolution of the front can be captured by
the time-dependence in Eq.~(\ref{genfisher}) with the same initial
conditions or if additional terms which include the time-dependent
interactions between the bacteria and the hostile environment are
necessary.

We thank Karin Dahmen, Nadav Shnerb, and David Nelson for many
useful discussions. We thank Jeremy Newburg-Rinn for help in
acquiring data, and Anna Delprato and Nancy King for helping us
with  technical aspects of culturing {\it Bacillus subtilis}.

\begin{figure} 
\begin{center} 
\end{center} 
\caption{(a) An image of a {\it Bacillus subtilis} colony growing
on nutrient rich agar at $t = 23.15$ hours (h) after point inoculation.
The region inside the
dashed line is shielded from UV radiation. (b) At a later time ($t =
65.45$ h), the circular front is distorted due to the
confinement of the bacteria to the shielded region.
(c) The width of the bacterial colony along the shield ($x$) and
perpendicular to the shield ($y$). The time-dependence of the
diameter $d$ of a colony growing in a petri dish which is completely
shielded from UV radiation is also shown. The horizontal
dotted line corresponds to the boundary of the shielded region, and
the vertical dotted line corresponds to the time when UV-radiation
is switched off. The temperature in all cases was
$21^\circ {\rm C} \pm 0.5^\circ {\rm C}$.}
\label{fisher} 
\end{figure} 

\begin{figure} 
\begin{center} 
\end{center} 
\caption{Images of the bacterial colony when the shielded region is
rotated at a constant angular velocity $\omega = 6.69
\times 10^{-6} s^{-1}$. (a) $t = 24$ h, (b) $t = 46.56$
h, (c) $t = 73.73$ h. The imposed convection 
velocity increases linearly from the axis of rotation. The
bacteria follows the shielded region at low values of $v$, but
lag behind far from the axis of rotation 
corresponding to higher velocities of the shield.}
\label{rot-pic} 
\end{figure} 

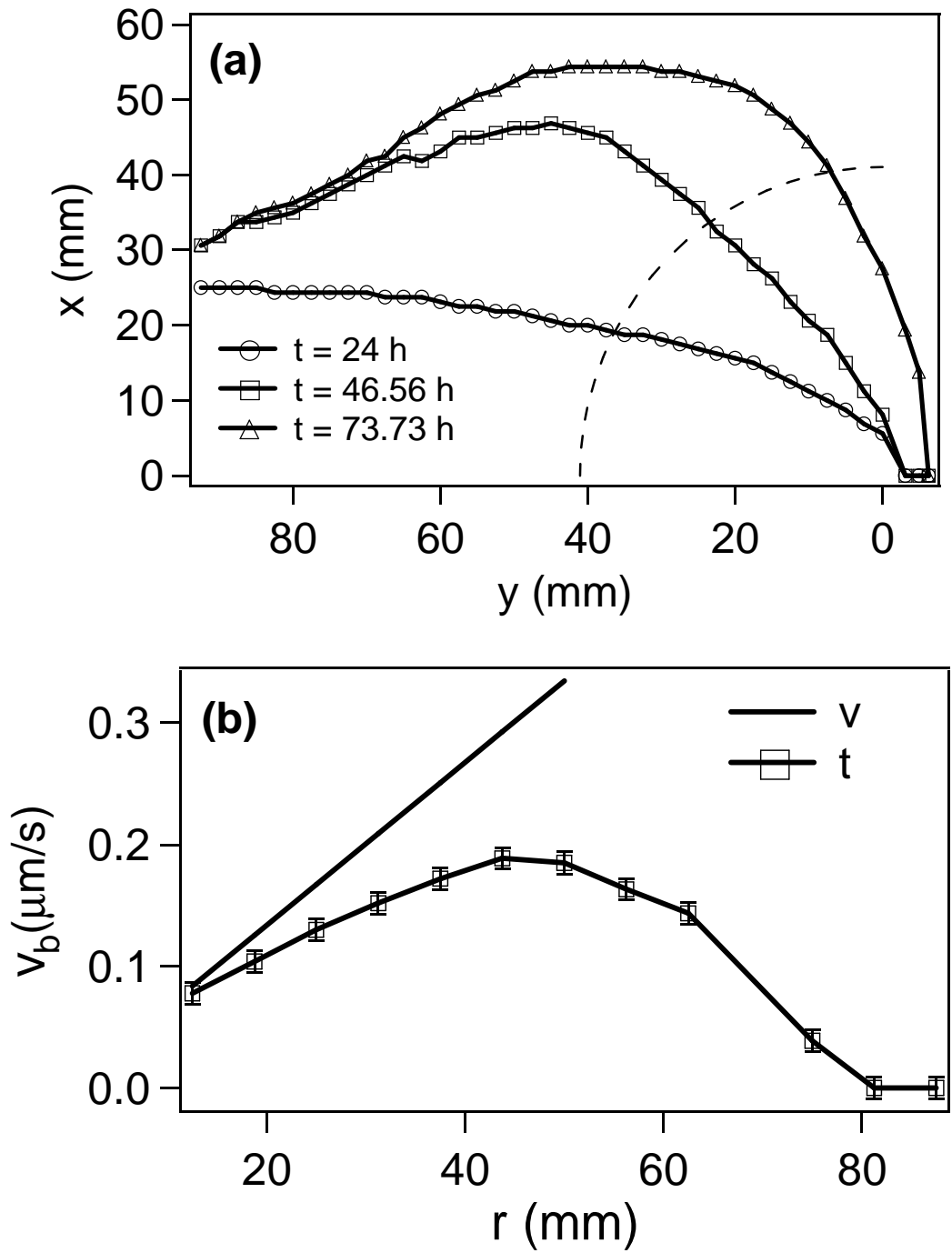
\begin{figure} 
\begin{center} 
\end{center} 
\caption{(a) The colony fronts extracted from the images shown in 
Fig.~\ref{rot-pic}. The axis of rotation is at (0,0) and the initial 
line of inoculation is along the horizontal axis. The distance where 
the speed of the shield corresponds to the Fisher wave velocity of 
the bacteria around $22^\circ$ is shown by the
dashed circle. (b) The 
speed of the bacteria front $v_{b}$ is observed to increase to
$0.23\,\mu {\rm m/s}$ and then decrease. This maximum speed
corresponds to $v_{F}~0.26\,\mu {\rm m/s}$ around 22$^\circ$C.} 
\label{front-pic}
\end{figure} 

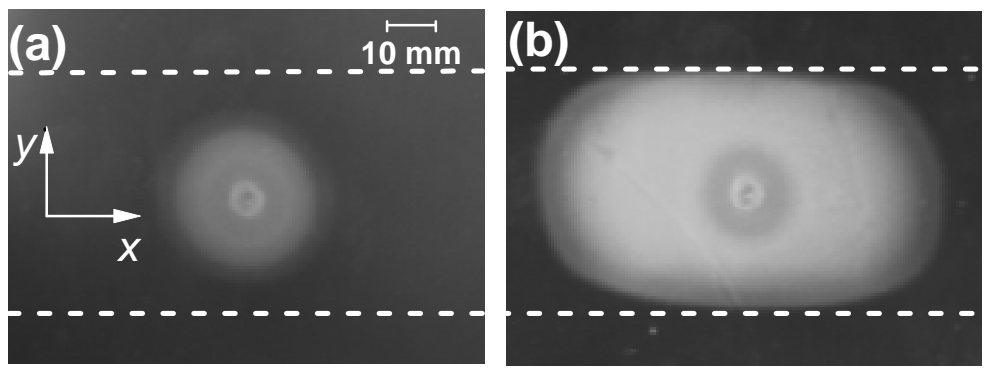
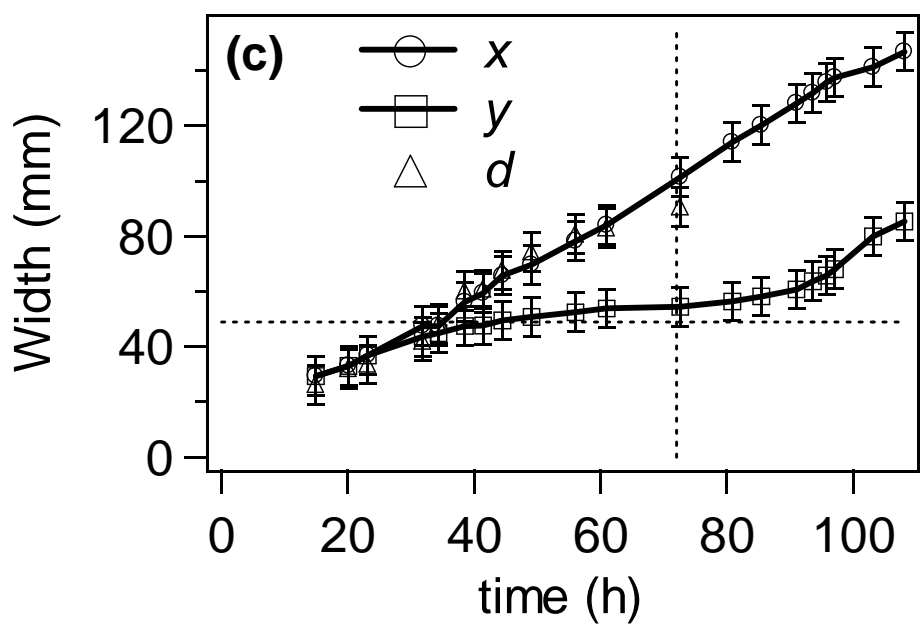
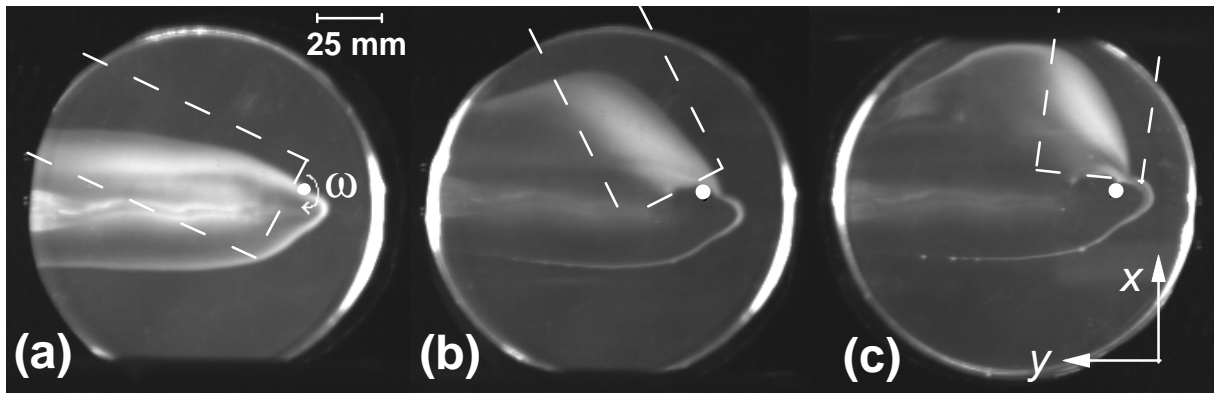
\end{document}